\RequirePackage[hyphens]{url}

\documentclass{article}
\usepackage{biblatex}
\usepackage{amsmath,amsfonts,amsthm}
\usepackage{xspace}
\usepackage{hyperref}
\usepackage{url}
\usepackage[hyphenbreaks]{breakurl}
\usepackage{tikz}

\usetikzlibrary{positioning}
\usetikzlibrary{decorations.pathreplacing}
\usetikzlibrary{arrows.meta}
\tikzset{>={Latex[width=1.5mm,length=1.5mm]}}

\newtheorem{theorem}{Theorem}

\newcommand{\content}{\ensuremath{\mathsf{m}}\xspace}
\newcommand{\recoveryRate}{\ensuremath{\mathsf{b}}}
\newcommand{\upperBound}{\ensuremath{\mathsf{M}}}
\newcommand{\Commit}{\ensuremath{\mathsf{Commit}}\xspace}

\newcommand{\commitment}{\ensuremath{\mathsf{com}}\xspace}
\newcommand{\opening}{\ensuremath{\mathsf{op}}\xspace}
\newcommand{\Encrypt}{\ensuremath{\mathsf{Enc}}\xspace}

\newcommand{\symmetricKey}{\ensuremath{\mathsf{K}}\xspace}
\newcommand{\ciphertext}{\ensuremath{\mathsf{ct}}\xspace}
\newcommand{\Sign}{\ensuremath{\mathsf{Sign}}\xspace}
\newcommand{\secretKey}{\ensuremath{\mathsf{sk}}\xspace}

\newcommand{\signature}{\ensuremath\sigma\xspace}
\newcommand{\handle}{\ensuremath{\mathsf{h}}\xspace}

\newcommand{\recipient}{\ensuremath{\mathsf{r}}}
\newcommand{\senderMarker}{\ensuremath{\mathsf{s}}}

\newcommand{\identity}{\ensuremath{\mathsf{ID}}\xspace}
\newcommand{\accountabilityServer}{\ensuremath{\mathsf{AS}}\xspace}
\newcommand{\reputation}{\ensuremath{\mathsf{y}}\xspace}
\newcommand{\token}{\ensuremath{\mathsf{t}}\xspace}
\newcommand{\timestamp}{\ensuremath\tau\xspace}
\newcommand{\bonus}{\ensuremath{\mathsf{k}}\xspace}

\newcommand{\reportCount}{\ensuremath{\mathsf{x}}\xspace}
\newcommand{\noise}{\ensuremath{\mathsf{N}}\xspace}
\newcommand{\noiseDistr}{\ensuremath\mathcal{N}\xspace}
\newcommand{\state}{\ensuremath{\mathsf{sc}}\xspace}
\newcommand{\probReport}{\ensuremath{\mathsf{p}}\xspace}
\newcommand{\probReward}{\ensuremath{\mathsf{q}}\xspace}
\newcommand{\reputationFunc}{\ensuremath{\mathsf{upd}}\xspace}
\newcommand{\rewardLabel}{\ensuremath{\mathsf{Reward}}\xspace}
\newcommand{\oursystem}{\ensuremath{\mathsf{Sandi}}\xspace}

\renewcommand{\paragraph}[1]{\vspace{.6em}\noindent\textbf{#1.}\hspace*{.5em}}
\renewcommand{\subparagraph}[1]{\vspace{.3em}\noindent\textit{#1.}\hspace*{.5em}}

\title{\oursystem: A System for Accountability and Applications in Direct Communication\\\vspace{1ex}\large Extended Abstract}

\author{
  F. Betül Durak\thanks{Microsoft Corporation} \and
  Kim Laine\thanks{Microsoft Corporation. Corresponding author: \href{mailto:kim.laine@microsoft.com}{\nolinkurl{kim.laine@microsoft.com}}} \and
  Simon Langowski\thanks{Massachusetts Institute of Technology} \and
  Radames Cruz Moreno\footnotemark[1] \and
  Robert Sim\footnotemark[1] \and
  Shrey Jain\footnotemark[1]
}

\date{November 2023}

\begin{document}

\maketitle

\begin{abstract}
Reputation systems guide our decision making both in life and work: which restaurant to eat at, which vendor to buy from, which software dependencies to use, and who or what to trust. These systems are often based on old ideas and are failing in the face of modern threats. Fraudsters have found ways to manipulate them, undermining their integrity and utility. Generative AI adds to the problem by enabling the creation of real-looking fake narratives at scale, creating a false sense of consensus. Meanwhile, the need for reliable reputation concepts is more important than ever, as wrong decisions lead to increasingly severe outcomes: wasted time, poor service, and a feeling of injustice at best, fraud, identity theft, and ransomware at worst.

In this extended abstract we introduce \oursystem, a new kind of reputation system with a single well-defined purpose: to create trust through accountability in one-to-one transactions.
Examples of such transactions include sending an email or making a purchase online.
\oursystem has strong security and privacy properties that make it suitable for use also in sensitive contexts. 
Furthermore, \oursystem can guarantee reputation integrity and transparency for its registered users.

As a primary application, we envision how \oursystem could counter fraud and abuse in direct communication.
Concretely, message senders request a cryptographic tag from \oursystem that they send along with their message.
If the receiver finds the message inappropriate, they can report the sender using this tag.
Notably, only senders need registered accounts and do not need to manage long-term keys.
The design of \oursystem ensures compatibility with any communication system that allows for small binary data transmission. 
\end{abstract}

\section{Introduction}\label{sec:intro}
The Internet was built without an identity layer~\cite{cameron2005laws}.
This results in inherent pseudonymity~\cite{van2017usernames}, which on one hand promotes freedom of speech and enables people to express their identities and views more openly.
On the other hand, it complicates the attribution of unsatisfactory or inappropriate online transactions to real-world people, organizations, and companies.
In other words, it is hard to hold people accountable.

Inability to do this is detrimental to online stores and marketplaces, messaging platforms, social media platforms, and any other platform that facilitates transactions.
The problem is amplified by scale~\cite{cerf2023dilemma}: we now need to choose from a vast number of sellers, business partners, or service providers. We need to choose which emails to open and which phone calls to answer.
Wrong decisions lead to increasingly severe outcomes: wasted time, poor service, and a feeling of injustice at best, fraud, identity theft, and ransomware at worst.
In this situation, how can any online platform possibly establish trust between its vast number of users?

\subsection{From Reputation to Accountability}
Reputation systems have been successfully used to address these problems in many scenarios~\cite{josang2007survey,hendrikx2015reputation,farmer2010building}.
Just a few examples demonstrating the breadth in deployment of reputation systems include likes and dislikes on social platforms, GitHub stars, ratings and reviews on Amazon, seller ratings on eBay, and karma on Reddit.
All these systems were designed with a variety of intentions -- some social, others to establish ``trustworthy'' accounts that are essential for the platform to provide value.
However, many reputation systems are based on decades old ideas and are starting to fail in the face of modern threats: fake reviews, review farms, reputation inflation, privacy problems, lack of transparency, revenge ratings, among many others.
Generative AI has further exacerbated these concerns, as it can create realistic looking fake narratives about people, products, and services.

In this paper, we introduce a system called \oursystem, a new kind of reputation system with a single well-defined purpose: to create trust through accountability in one-to-one transactions.
By one-to-one transactions we mean any transaction between two parties that may have asymmetrical roles, such as seller and buyer, or sender and receiver.
Independent of the nature of the transaction, will call these two parties a \textit{sender} and a \textit{receiver}, due to our leading application in direct communication that we get to in a moment (\autoref{sec:direct_comm}).

Traditionally, accountability (in one-to-one transactions) would be enforced through an authority that both participants answers to.
However, in \oursystem no one entity is in absolute power to hold others accountable.
The dynamic we create is carefully crafted to create no power imbalance, yet we can prove it achieves our desired goal: incentivizing transactions where the sender is less likely to get reported.
We believe \oursystem can enhance trust and help people make better decisions in all forms of one-to-one transactions, while addressing the shortcomings of traditional reputation systems in the face of modern threats.

In \oursystem, senders have reputation scores that the receivers can see.
During a transaction, the sender gives the receiver an \emph{endorsement tag}.
This is a kind of receipt that confirms the receiver has engaged in a transaction with the sender and gives the receiver permission to hold the sender accountable for an unsatisfactory transaction (poor quality service or product, poor customer support, inappropriate email).
The receiver can at will \textit{report} the sender to the system using the endorsement tag, thus reducing the sender's score.
This asymmetry makes the system more flexible and easier to adopt, as only the senders need to register accounts with the system, whereas the receivers do not.
The reputation system itself, including storing the scores for each sender, is handled by an \emph{accountability server} (\accountabilityServer).

Finally, \oursystem is practical, as it can be easily implemented on top of almost any kind of transaction method.
All it requires is senders and receivers to be able to communicate a few hundred bytes of data with \accountabilityServer (\autoref{sec:implementation}).
It also has strong security and privacy properties, which are needed in different applications depending on the sensitivity of the transactions.
Some of these properties are also crucial to protect \oursystem from manipulation (\autoref{sec:security_privacy}).

\subsection{Accountability in Direct Communication}\label{sec:direct_comm}
While \oursystem has many applications (we will discuss some more in \autoref{sec:discussion}), we will focus on a particular use-case to illustrate our thinking and our construction: \textit{direct communication}.

By direct communication, we mean a one-to-one transaction of information, where a sender sends a message to a receiver.
Examples of direct communication include email, messages on systems like Teams, Zoom, WhatsApp, and DMs on social media platforms. In many cases, the communication may be end-to-end encrypted.
Many communication platforms allow unlimited unsolicited messages but others require a basic acknowledgment, \textit{e.g.}, accepting a contact request.
Because some unsolicited messages can be important, such as job offers or bank alerts, it is crucial to be able to have at least some trust in unsolicited communication without prior knowledge of the sender (including no authentication).
On the other hand, generative AI can be misused to create believable fake narratives for fraud, scams, or political interference.
Identifying such fake narratives can be challenging; for example, we cannot simply detect AI generated content, since these models are also being used to enhance legitimate and benign communications.
This scenario is ideal for \oursystem, as it can bring trust in these inherently untrusted contexts.

Finally, we note that one-to-many communication does not fit our model; examples of such would be posting a message publicly on social media, or sending an email to a mailing list without knowing who the receivers are.
We explain the problems with one-to-many communication in \autoref{sec:discussion}.

\subsection{Sandi}\label{sec:sandi}
As we already explained above, \oursystem involves three entities: an accountability server \accountabilityServer, registered senders, and receivers.
For each sender account, \accountabilityServer maintains a score that \accountabilityServer decreases as it receives reports.
In this sense, the reputation system in \oursystem is monotone (there is no up-voting).
However, it also has a carefully crafted automatic recovery mechanism, which counters accidental or malicious reports, and eventually recovers the reputation for senders that have improved their behavior.

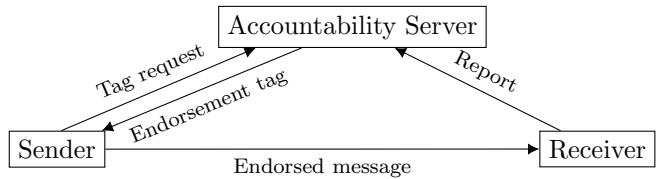
\begin{figure}
    \begin{center}
        \begin{tikzpicture}[node distance=3cm]
            % Nodes
            \node (score) [draw, rectangle] {Accountability Server};
            \node (sender) [draw, rectangle, below left=1.1cm and 1.5cm of score] {Sender};
            \node (recipient) [draw, rectangle, below right=1.1cm and 0.7cm of score] {Receiver};
            
            % Edges
            \draw[->] ([xshift=7mm]sender.north west) -- node[sloped, anchor=center, above] {\footnotesize Tag request} ([xshift=5mm]score.south west);
            \draw[->] (score) -- node[sloped, anchor=center, below] {\footnotesize Endorsement tag} (sender);
            \draw[->] (sender) -- node[below] {\footnotesize Endorsed message} (recipient);
            \draw[<-] (score) -- node[sloped, anchor=center, above] {\footnotesize Report} (recipient);
        \end{tikzpicture}
    \end{center}
    \caption{Data flow in $\oursystem$. A sender and \accountabilityServer run a protocol that results in the sender receiving an endorsement tag for their message. The tag is passed along with the message to a receiver. The receiver can optionally report the message by sending the endorsement tag back to \accountabilityServer.}
    \label{fig:info_flow}
\end{figure}

After registering an account, a sender interacts with \accountabilityServer to get a cryptographic \textit{endorsement tag} for its message.
This tag includes, among other things, the sender's current reputation digitally signed by~\accountabilityServer.
Upon receiving the message with the tag (we call this an \textit{endorsed message}), the receiver can verify that the tag corresponds to the message it received.
If it deems the content inappropriate, it reports the sender simply by sending the tag to~\accountabilityServer.
This simplicity makes \oursystem compatible with all existing communication systems that can support small binary data transmission.
The information flow (ignoring many details) between different parties is summarized in \autoref{fig:info_flow}.

In essence, receivers judge the appropriateness of messages, which forces senders to be cautious in their approach.
As an example, if a political campaign indiscriminately sends out messages (high precision and high recall), their reputation may drop rapidly, discouraging such practices. Even with \oursystem deployed, there is no requirement for anyone to send endorsed messages: the endorsement tag simply aims to make it more likely that the receiver will consider the message to be genuine and worth reading.
This thinking applies in other uses of \oursystem as well. An explicit endorsement (for the transaction) and a visible reputation simply make it easier to find trustworthy partners for any one-to-one transaction.

The senders' reputation is determined by a \textit{score function} that maps their current reputation and the number of reports they have received in a fixed time period to their new reputation.
A lower reputation makes it less likely for receivers to pay attention to their messages, thus lowering the senders' expected gain.
Indeed, in the full version of this paper, we show that \accountabilityServer can control the senders' optimal strategy to be such that inappropriate messages are sent arbitrarily far in the future. This idea is captured below in the informal \autoref{th:optimality_informal}.

\paragraph{Whitewashing} Whitewashing, where users easily quit and create new accounts, poses a challenge for reputation systems.
For systems like \oursystem that focus on accountability, whitewashing undermines the entire idea of being able to hold reputation owners accountable~\cite{mirkovic2008building}.

Several solutions to whitewashing have been proposed, including requiring proof of a strong identity~\cite{mirkovic2008building}, relying on an existing PKI~\cite{koutrouli2012taxonomy,xiao2016survey}, disadvantaging fresh accounts~\cite{friedman2001social}, and adding ``friction'' to the account creation process~\cite{koutrouli2012taxonomy,kozlov2020evaluating}, such as CAPTCHAs or phone-based verification.

Whitewashing is an independent problem that we are not trying to solve in \oursystem.
Thus, from now on we assume the account creation process is resistant to Sybil attacks, preventing a single malicious sender from creating multiple accounts.

\subsection{Related Work}
We compare \oursystem to prior work in several related areas: reputation systems, spam filtering techniques, and content moderation (message franking).

\paragraph{Reputation systems}
There is a vast amount of literature on reputation systems, including \cite{josang2007survey,hendrikx2015reputation,farmer2010building,steinbrecher2006design,kerschbaum2009verifiable,goodrich2011privacy,clauss2013k,schiffner2011limits,blomer2015anonymous,schaub2016trustless,zhai2016anonrep,bemmann2018fully,blomer2018practical,bag2018privacy,hauser2023street}.
Comprehensive surveys of ``privacy-preserving'' reputations systems can be found in \cite{gurtler2021sok,hasan2022privacy}.
Such reputation systems tend to provide various forms of privacy, security, integrity, and transparency guarantees.
Schiffner \textit{et al.}~\cite{schiffner2011limits} and Clau{\ss} \textit{et al.}~\cite{clauss2013k} present formal privacy definitions for reputation systems, but their setting is incomparable to ours.
Hauser \emph{et al.}~\cite{hauser2023street} explore the problem of aggregating reputation scores from multiple sources in a secure and private manner.
In a recent paper, Bader \textit{et al.}~\cite{bader2023reputation} discuss the challenges of using reputation systems for business-to-business transactions and outline a proposal based on fully homomorphic encryption. We discuss this topic later in \autoref{sec:other_uses}.

Our hypothesis is that privacy technologies alone cannot address the problems reputation systems face today~\cite{sekar2023technical,he2022market,movschovitz2013analysis,resnick2002trust}.
This is why with \oursystem we take a more comprehensive approach. We focus on a narrow well-defined scenario where we can clearly express what the purpose of the system is, utilize modern privacy technologies, and use a sophisticated score function that we can argue provably drives participants towards the desired kind of behavior.
In short, the reputation system underlying \oursystem is a privacy-preserving centralized, \textit{unidirectional} (only receivers can report senders), and ``somewhat monotone'' (reports always indicate a negative rating, but we include a fixed recovery rate). The adversarial model and our security and privacy guarantees are listed in \autoref{sec:security_privacy}.

Denial-of-Service (DoS) type attacks against reputation systems~\cite{hoffman2009survey} are difficult to protect again, and we leave them outside our threat model.
For example, \accountabilityServer could simply refuse to respond to messages from senders or from receivers.

\paragraph{Spam filters}
Ever since the mid-1990s, spam has impacted nearly all popular online communication systems~\cite{krebs2014spam,cormack2008email,grier2010spam}.
Various technical and legal anti-spam solutions have been designed and deployed over the years~\cite{ramachandran2007filtering,grier2010spam,esquivel2010effectiveness,chirita2005mailrank,prakash2005fighting,zheleva2008trusting,zhang2009ipgrouprep,guzella2009review,dada2019machine}.
Authentication and anti-spoofing systems~\cite{IETF-SPF,IETF-DKIM,IETF-DMARC} can help, but a recent study~\cite{wang2022large} found numerous problems with their adoption.
Others have considered proof-of-work style approaches, where senders spend some resource to send messages~\cite{pricing_via_combatting_junk_mail,pow_spam_filter,liu2006proof,pow_cant_work,pow_cant_can_does_work}, but these unfairly disadvantage people who legitimately need to send more.

In contrast, \oursystem would not aim to block any messages, but provides signal of trust through accountability. 
Senders can view their score, which is not the case with spam filters.
The receiver's view is similar to ``report spam'' buttons, but a report in \oursystem has much more direct impact.
\oursystem only penalizes bad behavior, unlike proof-of-work systems, where everyone pays.

\paragraph{Content moderation and message franking}
Content moderation~\cite{kamara2022outside} is typically implemented within tightly controlled systems, whereas \oursystem places no limitations on the transaction (or communication) system.
In end-to-end encrypted communication the primary interest shifts to receiver-based reporting of inappropriate messages, which is also the approach we take: \oursystem never has visibility into any message, whether it is reported or not.

As an example, in the \emph{message franking} technique~\cite{kamara2022outside,meta_message_franking,tyagi2019asymmetric,issa2022hecate} a cryptographic tag, comprising sender and receiver names, message commitment, and timestamps, is included in metadata for an end-to-end encrypted message.
This message can then be is passed around from user to user and the metadata is updated accordingly.
Once a receiver reports it, the message and its tag are revealed to the system.

Despite some similarities, \oursystem and message franking are fundamentally different. \oursystem is geared towards building trust through accountability, whereas the primary objective of message franking is to aid content moderators in their evaluations and decisions.

% \fi

\section{Security and Privacy} \label{sec:security_privacy}
\oursystem needs to provide strong security and privacy properties to earn trust and avoid causing unintended harms for its users.
We assume that \accountabilityServer does not collude with any sender, because they can together break a receiver's privacy and inflate the sender's score arbitrarily.
Extensions of \oursystem to mitigate these issues will aim to present solutions in our future work. 

Informally, we have the following adversarial model and security and privacy objectives \oursystem provides:
\begin{itemize}
\item There are a few ways an honest sender can become a victim of slandering. 
(1) A malicious \accountabilityServer can try to incorrectly lower an honest sender's score without being caught. 
(2) \emph{A malicious receiver} can try to report an honest sender more times than the number of messages it receives from the honest sender.
(3) A malicious sender can try to create endorsement tags associated with another honest sender.

\noindent
\textsf{[Score Integrity]} We address these issues all together as follows.
One of the senders, the victim, is honest, while the rest of the senders, \accountabilityServer, and some subset of receivers are all malicious. 
Let \textsf{s} be the number of messages the honest sender sends. 
Some of these messages would be sent to malicious receivers and some to honest receivers: $\textsf{s} = \textsf{s}_{\textsf{malicious}} + \textsf{s}_{\textsf{honest}}$. 
Let $\textsf{r}_{\textsf{honest}}$ be the number of reports from honest receivers. 
Without loss of generality, we assume that all messages sent to malicious receivers are reported. 
The adversary's goal is to make an honest sender accept that the total number of reports it received is greater than $\textsf{r}_{\textsf{honest}} + \textsf{s}_{\textsf{malicious}}$.
In the full version of this paper we prove that this is impossible.
Note that it can never make the sender accept that the number of reports is greater than $\textsf{s}_{\textsf{honest}} + \textsf{s}_{\textsf{malicious}}$, because the sender knows it never sent that many endorsed messages.

\item  A malicious \accountabilityServer can try to find who an honest sender communicates with.

\noindent
\textsf{[Communication Privacy]} If an honest sender sends a message to an honest receiver, a malicious \accountabilityServer cannot learn any information about the message nor to which receiver the message is sent. 
However, \accountabilityServer learns (a) whether the message is reported and (b) information about the reporter that is leaked by network traffic, unless reporting is done through anonymous routing.

\item A malicious sender can try to detect if a receiver reported the sender to an honest \accountabilityServer. 

\noindent
\textsf{[Report Privacy]} This notion is captured in two phases. In the full version of this paper we show that (a) a malicious sender cannot distinguish a reporter from a non-reporter if it controls neither, and (b) the true number of reports is hidden due to a differential privacy guarantee.

\item A malicious receiver can try to distinguish if two distinct sender addresses belong to the same owner based on their endorsement tags.

\noindent
\textsf{[Unlinkability]} A malicious receiver has no advantage in guessing whether two endorsed messages come from two different honest senders or the same sender, beyond what can be inferred from the message contents, timestamps, network traffic, and sender reputations, without colluding with a malicious \accountabilityServer.
\end{itemize}

\section{How \oursystem Works}\label{sec:overview-building-up}
In this section, we explain at a high level how \oursystem works.
We provide a description of the most basic version of \oursystem with weaker security and privacy guarantees.
In the full version of this paper we show how to extend \oursystem to cover also the remaining guarantees.
The parties involved are the accountability server \accountabilityServer, a set of senders, and a set of receivers.

\paragraph{Setup}
Each sender must register with \accountabilityServer, which sets up a database entry for them indexed by an internal account identity~$\identity_\senderMarker$.
This entry contains authentication details, the sender's \textit{score} \state, and a report count \reportCount initialized to zero.
Both \accountabilityServer and the sender can view \state, whereas receivers see a coarsened view we call the sender's \textit{reputation}, denoted~$\reputation_\senderMarker$.
The two different views are used to improve the sender's privacy, ensuring a receiver cannot link two sender addresses to a single account simply by guessing based on matching reputations.
The reputation values are in a totally ordered set, for example \{``very high'', ``high'', ``medium'', ``low''\}, whereas the score is an integer or a real number.
The score is mapped to the reputation through a monotonically increasing \textit{reputation function} $\reputation(\cdot)$.

Time in \oursystem proceeds in epochs, the duration of which can be freely configured by \accountabilityServer (\textit{e.g.}, 1 day, 2 days, 7 days).
During each epoch, a set of messages are sent by the senders to their desired receivers.
At the end of the epoch, the score is updated using a \textit{score function \reputationFunc} that maps the sender's current score and the current number of reports to the new score for the next epoch.
The report counter \reportCount is reset to zero.
The choice of the score function is of utmost importance. We discuss this very briefly below in \autoref{sec:scoring} and present a full analysis in the full version of this paper.

\paragraph{Endorsed messages}
When a sender wants to send a message \content to a recipient, it needs to first authenticate with \accountabilityServer, so that \accountabilityServer knows the sender's account $\identity_\senderMarker$.
Next, it computes a cryptographic commitment (and opening) to its message and the reciever, as $(\commitment, \opening) \gets \Commit(\content, \handle_{\recipient})$, and sends $\commitment$ to~\accountabilityServer.
Subsequently, \accountabilityServer encrypts the sender's $\identity_\senderMarker$ with a secret encryption key \symmetricKey as $\ciphertext \gets \Encrypt_\symmetricKey(\identity_\senderMarker)$ and uses a secret signing key \secretKey to produce a digital signature $\signature \gets \Sign_{\secretKey}(\commitment||\timestamp||\reputation_\senderMarker||\ciphertext)$.
Here \timestamp is a timestamp and $\reputation_\senderMarker$ the sender's reputation.
It creates an \emph{endorsement tag} $\token \gets (\commitment, \timestamp, \reputation_\senderMarker, \ciphertext, \signature)$ which it sends to the sender.

Once the sender has received the endorsement tag, it verifies that the signature is valid using \accountabilityServer's public verification key.
It can then send the tuple $(\token, \content, \opening)$ to the receiver.
We call this tuple an \emph{endorsed message}.

Upon receiving and parsing $(\token, \content, \opening)$, the receiver verifies that $(\commitment, \opening, \content)$ and \signature are valid.
If the check passes, it uses $\reputation_\senderMarker$ to decide whether it wants to further engage with the content, \emph{e.g.}, read it or react to it.
If the receiver considers \content to be inappropriate, it sends the endorsement tag \token to \accountabilityServer.
We call this a \emph{report}.

Upon receiving and parsing a report \token, \accountabilityServer verifies \signature, decrypts \ciphertext to obtain $\identity_\senderMarker$ and increments the sender's report count~\reportCount.
When the epoch changes, \accountabilityServer looks at \reportCount for the sender and updates $\state$ according to the score function as $\state \gets \reputationFunc(\state, \reportCount)$.

\paragraph{Improvements}
The above sketch omits several issues that we will address in the full version of this paper.
Namely, of the security and privacy properties discussed above in \autoref{sec:security_privacy} only \textsf{Communication Privacy} and \textsf{Unlinkability} are satisfied.

For example, in what we described above, \accountabilityServer could just change each sender's scores as it pleases.
In some scenarios this may not be a problem and \accountabilityServer can be trusted to behave correctly, but in other scenarios there may be concerns that \accountabilityServer misbehaves. 
Another issue is that the privacy of reports is not protected.
Namely, it would be easy for a sender to learn whether a particular (possibly the only) endorsed message it sent out was reported.
This is problematic, for example, if someone wants to report an inappropriate message from their colleague or business partner.
In the full version of this paper we show how to extend \oursystem to address these shortcomings to satisfy also \textsf{Score Integrity} and \textsf{Report Privacy}.

\section{Scoring Mechanism}\label{sec:scoring}
A core part of any reputation function is a \textit{score function} that explains how reputation is computed.
Common choices are averages and (weighted) sums, but also more exotic functions have been used.
Whichever function is used, one should be able to argue why it results in a notion of reputation that drives participants towards some desired behavior (buy less from shady vendors, use higher quality software dependencies, \emph{etc.}).

In each epoch, a set of messages are sent by the senders to their desired receivers.
Upon receiving a message \content from a sender with reputation $\reputation_\senderMarker$, receivers can choose to (not exclusively)
\begin{enumerate}\setlength\itemsep{0ex}
\item Reward the sender for the message with probability $\probReward(\content, \reputation_\senderMarker)$, \textit{i.e.}, read, engage with, or respond to the message in some way that creates often a mutual benefit for the sender and receiver, measured by a function $\rewardLabel(\content)$. The expected reward is $\probReward(\content, \reputation_\senderMarker) \rewardLabel(\content)$.
\item Report the message as inappropriate with probability $\probReport(\content, \reputation_\senderMarker)$, reducing the sender's score, and subsequently reputation, for the next epochs.
\end{enumerate}

For a fixed \content, we assume $\probReward(\content, \cdot)$ is increasing and $\probReport(\content, \cdot)$ decreasing.
These functions can result in a reward (resp. a report) even if $\content$ is inappropriate (resp. appropriate).
Naturally, $\probReward$ and $\probReport$ depend on the sender's reputation $\reputation_\senderMarker$ that the receiver sees, just as warning banners decrease click rate~\cite{hu2018end}.
As $(\probReward, \probReport)$ is in reality specific to a receiver, targeting receivers is equivalent to choosing $(\probReward, \probReport)$ among available pairs.
This marketing problem is out of scope for our construction, so we assume that the choice is made and that $(\probReward, \probReport)$ is the same for all receivers.

\paragraph{Our score function}
We use the following score function:
\begin{equation}
  \reputationFunc_{\bonus, \recoveryRate}^\upperBound(\state, \reportCount) :=
    \begin{cases}
      \state - \reportCount + \bonus~ \text{if } \reportCount \geq \bonus \\
      \min  \left\{ \state + \recoveryRate, \upperBound \right \} ~ \text{if } \reportCount < \bonus,\,\state \geq 0\\
      \min \left\{ \state - \reportCount + \bonus, 0 \right \} ~ \text{if } \reportCount < \bonus,\, \state < 0\\
    \end{cases}
    \label{eq:our_upd}
\end{equation}
The function takes as inputs the sender's current score \state and their report count (in this epoch)~$\reportCount$.
The parameter $\upperBound$ is an upper bound on the score; there is no lower bound.
The parameter $\bonus \geq 1$ is a \emph{tolerance level}; if the sender receives fewer than \bonus reports within an epoch, their reputation will not be affected.
Finally, the parameter $\recoveryRate \in (0, 1]$ is a real number that determines how much the score recovers per epoch if fewer than \bonus reports are received.

The function in \autoref{eq:our_upd} looks very unusual and is unlike anything used in reputation systems before.
However, in the full version of this paper we prove that it satisfies a list of natural properties we would expect from such a score function.
Furthermore, we prove that \textit{any} score function satisfying these properties leads to the following result that governs the behavior for ``logical'' senders.
\begin{theorem}[informal]
    Given a score update function $\reputationFunc_{\bonus,\recoveryRate}^\upperBound$ and a reputation function $\reputation$, with $n$ epochs left in the game and sender's current score $\state$, there exists an optimal sender's strategy\footnote{An optimal strategy maximizes the sender's expected total reward.} where, in each epoch, the sender sends messages that maximize $\frac{\probReward(\content, \reputation_\senderMarker)}{\probReport(\content, \reputation_\senderMarker)} \rewardLabel(\content)$ until it has received a given number of reports which is no larger than ~$\bonus$, after which it waits for the next epoch.
\label{th:optimality_informal}
\end{theorem}

In some applications it is necessary to protect the privacy of reports.
This means that the sender would not know whether any particular message it sent out was reported.
To do this, \accountabilityServer applies differential privacy by first sampling $\noise \gets \noiseDistr$, where $\noiseDistr$ denotes some noise distribution, and then computes the updated score as $\state \gets \reputationFunc(\state, \reportCount + \noise)$.
With differential privacy, the optimality theorem changes somewhat, as we discuss in detail in the full version of this paper.

\section{Discussion}\label{sec:discussion}

\subsection{Incentives for Reporting}\label{sec:incentives}
The success of \oursystem is dependent on accurate and timely reports, otherwise the system loses its meaning.
This is a real threat already for all or most reputation systems.
For example, in \cite{fradkin2023incentives} Fradkin and Holtz analyze how the lack of incentives to review undermines the reputation system used by Airbnb.
Generally, the subset of people who provide feedback (reports, reviews, \textit{etc.}) tends to be biased.

Various real-world application mechanisms exist to incentivize users to submit reports.
For example, some messaging systems connect spam reporting with sender blocking, providing a recipient with a concrete benefit from filing a report.
This is comparable to \oursystem, where a report has a concrete effect on the sender's reputation.

Other systems, such as the Stack Exchange reputation system~\cite{movschovitz2013analysis}, intentionally gamify reputation so that users have incentives for filing both positive and negative reports, and filing these reports contributes to their own reputation.
In these systems, care must be taken not to incentivize users to file spurious reports~\cite{DBLP:journals/corr/abs-2111-07101}.
One of the goals of \oursystem is to instead eliminate gamification, as it is hard to know ahead of time what an optimal strategy ends up being in such a complex system.

Some reputation systems rely on implicit reporting, for example, by measuring (or predicting) the fraction of a sender's messages that are left unread or deleted, or considering other meta-features such as statistical message properties, relay server reputation, or social graph distance~\cite{4268083,8573330,US8370930B2}.
It should be noted that these mechanisms exist largely at the application layer and would need to be carefully integrated with the reputation system in order to be useful.
However, with sufficiently strong signals from application layer software, implicit reporting is fully compatible with \oursystem.

\subsection{One-to-Many Transactions}
\oursystem is intended only for one-to-one transactions.
To understand what the problem with one-to-many transactions is, consider the case of a public post to social media.
The problem is: \textit{who is supposed to be able to report an inappropriate message in this case?}

If anyone who sees the post can, then the sender may have just staked their entire reputation and may easily end up getting a massive number of reports.
Furthermore, now it would be impossible to prevent repeated reporting without requiring receiver accounts and proofs of a strong identity.
Thus, in this case a sender cannot be incentivized to participate at all.

On the other extreme, if the message can be reported only once (\accountabilityServer can enforce this), should the message even be shown as a valid endorsed message to the receivers?
We argue no, because each individual receiver's relative opportunity to hold the sender accountable for an inappropriate message is $1/\textsf{number\_of\_receivers}$, which can be arbitrarily close to zero.
Thus, the sender is not putting enough at stake to earn the added visibility from a reputed message.

It seems that there are no  meaningful alternatives between these two that would not require receivers to have accounts. If receivers had accounts, the \accountabilityServer could limit the number of reports to a fixed limit, perhaps chosen by the sender, and prevent multiple reports from each receiver.

\subsection{Positive Ratings}
For \oursystem, we chose to use a design that relies on negative ratings only (reports).
In principle, one might consider a system where the sender's reputation increases explicitly through a mechanism for positive ratings.
Note that in \oursystem positive ratings are implicit: few or no reports result in the sender's score increasing, according to \eqref{eq:our_upd}.
This choice was natural to us, as we identified several problems with explicit positive ratings.

Explicit positive ratings require receivers to be identified to prevent self-promotion, but this would violate \textsf{Communication Privacy} (\autoref{sec:security_privacy}) and reveal the ``trust graph'' to~\accountabilityServer.
Moreover, simply identifying the receivers is not enough: it would need to be possible to take down groups of adversarial accounts without allowing them to simply create new accounts, for example, by requiring proofs of a strong unique identity.
This is in stark conflict with the goals of \oursystem.

Another problem is that with only positive ratings there would be no cost or penalty in sending as many messages as possible.
Sending a message with no cost to get an engagement with some probability enables the strategy that advertisers (and spammers) already use today: send as many messages as you can to as wide of an audience as possible.
This creates a high precision-high recall situation, but since high recall has a low cost, this ends up being a good strategy.
The goal of \oursystem is not to prevent email advertisement campaigns, but to help people better understand the nature of the communication they receive.

One possible way of rewarding consistently high reputation or auxiliary evidence of good behavior is by changing the score function accordingly.
For example, one could increase \upperBound, increase \recoveryRate, or increase~\bonus. Each of these is beneficial to the sender and can be thought of as giving them more benefit of the doubt.

\subsection{Other Applications}\label{sec:other_uses}
\paragraph{Business-to-business transactions and supply chains}
A system like \oursystem can improve the trust and accountability in digital business transactions, which tend to be more unpredictable than in the past, when business relationships and trust were based on years of interaction and (successful) transactions.
A particularly interesting challenge lies in establishing trust in (software and other) supply chains that have become more volatile because of rapid development in technology, new logistic constraints, and geopolitical tensions, among other reasons~\cite{bader2023reputation}.

\oursystem can also help businesses make more informed choices.
For example, it could maintain reputation scores for businesses that reflect their reliability: high scores would indicate consistent and satisfactory results, and lower scores inability to meet expectations.
It would be possible to maintain multiple scores for different dimensions of reliability as well, \textit{e.g.}, product quality and customer support.

\paragraph{Business-to-consumer transactions}
Similarly, consumers today enjoy a vast number of choices for everyday transactions, such as buying groceries, home improvement, health services, or travel.
While there are multiple consumer facing reputation systems to help in making these choices, they suffer from unreliable (biased, misleading, fake) reviews~\cite{narciso2022unreliability,zheng2021identifying}.

Since \oursystem measures only unary negative signals, it cannot replace free-text reviews.
Nevertheless, the negative signal it provides can be particularly valuable for risk-averse consumers when making impactful purchases~\cite{lee2008effect,maheswaran1990influence}.
A 2016 study \cite{maharani2016discovering} found that people overwhelmingly prefer the 5-star rating system, hence it might be beneficial to map the senders' scores to a 1--5 domain with the reputation function~\reputation (\autoref{sec:overview-building-up}).

\subsection{Implementation}\label{sec:implementation}
We have implemented a prototype of \oursystem in Rust. We ran benchmarks in Azure on a standard \texttt{E16\_ads\_V5} VM (AMD EPYC 7763 @2.445 GHz). The benchmarks ran on a single thread.
To create an endorsement tag, the sender needs to communicate with \accountabilityServer and both need to do some computational work.
The total time (excluding networking) is $336\,\mu$s, of which \accountabilityServer's part is $150\,\mu$s.
The time it takes for \accountabilityServer to process a single report is $269\,\mu$s.

The serialized data communicated from \accountabilityServer to the sender in the tag issuance protocol is 236 bytes. The full endorsement tag from the sender to the receiver is 372 bytes.

\section{Conclusions and Open Problems}
We have presented a novel reputation system, \oursystem, with strong privacy and security properties.
\oursystem was constructed for a single well-defined purpose: to create trust through accountability in one-to-one transactions.
It addresses the shortcomings of traditional reputation systems by preventing the emergence of undesirable optimal strategies for their users and eliminates many of the modern threats these systems face.

Our contributions are twofold: (1) a detailed construction of \oursystem and its security justifications, and (2) an analysis of senders' behavior in \oursystem, with a proof that the system parameters can be used to effectively encourage users towards desirable behavior.
While \oursystem can help bring accountability and trust in any scenario with one-to-one transactions, our leading application was in countering inappropriate direct online communication.

There are many avenues for future work.
First, an exploration of different applications of \oursystem is needed.
Considering the ubiquity of reputation systems, determining the contexts where \oursystem provides the greatest value remains an open question.

Second, we already touched upon the challenge of incentivizing receivers to report in \autoref{sec:incentives}.
One idea to improve participation would be to provide concrete evidence to receivers that their reports had a tangible effect, but this is not included in the current construction.

Third, in some applications of \oursystem a lower score (or a report) may indicate a very serious problem, whereas in other applications it may indicate a minor nuisance. How the system parameters should be set to accommodate these different scenarios requires further study.

Fourth, it is crucial to validate whether \oursystem's improvements are recognizable and valued by users through user studies.
Assessing if it truly results in better decision-making compared to existing systems, and exploring its societal benefits from widespread adoption, are equally important.

\printbibliography

\end{document}